%% file: paper.tex
\documentclass[sigconf]{acmart}
\copyrightyear{2024}
\acmYear{2024}
\setcopyright{acmlicensed}\acmConference[MSR '24]{21st International
Conference on Mining Software Repositories}{April 15--16, 2024}{Lisbon,
Portugal}
\acmBooktitle{21st International Conference on Mining Software Repositories
(MSR '24), April 15--16, 2024, Lisbon, Portugal}
\acmDOI{10.1145/3643991.3645073}
\acmISBN{979-8-4007-0587-8/24/04}

\usepackage{nicefrac}

\usepackage[inline]{enumitem}

\newcommand{\DC}{$D_{C}$}
\newcommand{\DI}{$D_{I}$}
\newcommand{\DP}{$D_{P}$}

\newcommand{\CI}[1]{95\% CI$#1$}

\begin{document}

\title{How I Learned to Stop Worrying and Love ChatGPT}

\author{Piotr Przymus}
\orcid{0000-0001-9548-2388}
\email{piotr.przymus@mat.umk.pl}
\affiliation{%
  \institution{Nicolaus Copernicus University in Toruń}
  \country{Poland}}

\author{Mikołaj Fejzer}
 \orcid{0000-0003-1496-2289}
\email{mfejzer@mat.umk.pl}
\affiliation{%
  \institution{Nicolaus Copernicus University in Toruń}
  \country{Poland}}

\author{Jakub Narębski}
\orcid{0000-0002-3296-3915}
\email{jakub.narebski@mat.umk.pl}
\affiliation{%
  \institution{Nicolaus Copernicus University in Toruń}
  \country{Poland}}

\author{Krzysztof Stencel}
\orcid{0000-0001-6356-4872}
\email{stencel@mimuw.edu.pl}
\affiliation{%
  \institution{University of Warsaw}
  \country{Poland}}

\renewcommand{\shortauthors}{Przymus et al.}

\begin{abstract}
In the dynamic landscape of software engineering, the emergence of ChatGPT-generated code signifies a distinctive and evolving paradigm in development practices.
We delve into the impact of interactions with ChatGPT on the software development process, specifically analysing its influence on source code changes.
Our emphasis lies in aligning code with ChatGPT conversations, separately analysing the user-provided context of the code and the extent to which the resulting code has been influenced by ChatGPT.
Additionally, employing survival analysis techniques, we examine the longevity of ChatGPT-generated code segments in comparison to lines written traditionally.
The goal is to provide valuable insights into the transformative role of ChatGPT in software development, illuminating its implications for code evolution and sustainability within the ecosystem.
\end{abstract}



\maketitle

\input{1-introduction}
\input{2-methods}

\input{3-results}
\input{4-discussion}
\input{5-related}

\input{6-conclusion}

\bibliographystyle{ACM-Reference-Format}
\bibliography{paper.bib}

\end{document}

%% file: 1-introduction.tex
\section{Introduction}\label{introduction}

The rise of large language models (LLMs), like ChatGPT, is changing the way we do coding.
These AI tools can take on repetitive tasks, diving into the coding territory that used to be restricted for humans. 
While it got people pretty excited, there are also numerous concerns about the quality, security, and ethics of the code that comes out of ChatGPT (and similar tools, like GitHub Copilot).

One of the main concerns is the chance that AI-made code can bring in bugs and errors. 
These models lack the depth of understanding and expertise to fully grasp the nuances of complex code structures in existing projects.
This can lead to the generation of code with undetected errors, which can potentially cause software malfunctions and security vulnerabilities~\cite{DBLP:conf/msr/NguyenN22}.
Furthermore, such models may be used by inexperienced developers to learn general programming or explore available APIs~\cite{YILMAZ2023100005},
in which cases programmers may add subtle errors to existing codebase without understanding the impact of changes.

Additionally, there are privacy concerns, such as the inadvertent leakage of confidential code by developers to external entities.
They may further exploit such information during the training process~\cite{jaworski2023study}.
Moreover, there are concerns related to potential infringements on existing copyrights.
Since LLMs are trained on extensive amounts of publicly available text and code, there is a risk of generating code that violates copyrights or uses otherwise restricted intellectual property.

Despite these concerns, ChatGPT can prove to be a valuable tool when used responsibly.
For seasoned developers, ChatGPT can accelerate code generation, facilitating faster prototyping and experimentation~\cite{DBLP:journals/corr/abs-2305-11837}.
Nevertheless, it is crucial to thoroughly examine and analyze the code generated by ChatGPT before integrating it into production environments.

Acknowledging the inevitability of the use of code generated by LLMs, it is worthwhile to shift our focus from worrying about potential problems and instead embrace ChatGPT. 
Thus in this paper we try to asses what is the overall impact of ChatGPT on maintenance of software projects based on the MSR'24 challenge dataset~\cite{devgpt}.
In the following section, we strive to answer the following research questions.

{\bf RQ1.} To what extent does the usage of ChatGPT vary across different contexts of application, such as a commit, a pull request, or an issue?

{\bf RQ2.} What happens with code influenced by ChatGPT in the repository? How does its lifetime compare to similar lines created by human developers?

Replication package containing all custom tools and scripts developed for this paper is publicly available at Figshare~\cite{replication-package}.

%% file: 2-methods.tex
\section{Methods}
\label{sec:methods}

\subsection{Dataset}%
\label{sub:Dataset}
In this study, we analyzed the data from DevGPT dataset~\cite{devgpt}, which consisted of links to ChatGPT conversations in software repositories.
To analyze generated code we focused on merged changes related to \DP\ - pull requests, \DC\ - commits and \DI\ - issues (taken from latest sharing snapshot), excluding conversations with expired links (see Tab.~1. for details).
To enrich DevGPT dataset we utilize secondary data obtained from GitHub API and cloned Git repositories (see Tab.~2).
A specific repository can be referenced separately in \DC, \DP\ and \DI\ in different context. 
Then we established parent commits for each change from DevGPT files.
This varies depending on type of changes. 
It is explicit for changes in commits.
However, it required additional steps for pull requests (we computed diff for the whole pull request) and issues (we identified commits and pull requests closing the issue).
Following this, we determined whether the proposed changes were merged into the repository. The process was straightforward for pull requests, whereas for commits, we verified their presence in the main branch. Regarding issues, we assessed whether the associated pull request or commit had been merged, applying the same criteria as outlined above.

\subsection{Comparing ChatGPT conversations to code}%
\label{sub:Comparing ChatGPT conversations to code changes}
To calculate similarity between chat and code change we use Gestalt pattern matching algorithm (also known as Ratcliff and Obershelp algorithm\cite{ratcliff1988pattern}).
This algorithm is implemented in Python difflib standard library~\cite{10.5555/1593511,pythondifflib} and was used before in the context of code comparison~\cite{tsikerdekisPersistentCodeContribution2018}.
The similarity $0 \leq D_{ro} \leq 1$ of two strings $S_1$ and $S_2$ is determined by the formula
$D_{ro} = \frac{2K_m}{|S_1|+|S_2|}$ where $K_m$ is the number of matching characters.
The matching characters are defined as the longest common substring plus recursively the number of matching characters in the non-matching regions on both sides of the longest common substring.

For each commit and its corresponding conversation with ChatGPT, we assess the extent of information provided to and received from the chat. 
To achieve this, we compare the preimage of the commit, including its context (lines before the change), with the conversation prompts and potential snippets within it.
Similarly, we compare the postimage of the commit (lines after the change) with the conversation answers and potential snippets within it.
For every data hunk in a diff image, we search for the most suitable prompt/answer and code listing. Subsequently, we align each line in the hunk with lines in the corresponding prompt/answer, selecting lines with a similarity of at least 0.6 as recommended by \cite{10.5555/1593511,pythondifflib}.
All lines above this threshold are labeled as inspired by/inspiring ChatGPT and undergo further analysis.
Specifically, we examine the extent to which a commit is influenced by/influences ChatGPT and its duration of survival.

\subsection{Line survival analysis} 
Survival analysis \cite{liu2012survival} is useful in investigation of time-to-event data, like mechanical component failure. 
The \emph{survival function} $S(t)$ determines the probability that an event has not occurred up to time $t$, where $S(t) = 1 - F(t)$ with $F$ being the cumulative distribution function.
We utilize the Kaplan-Meier estimator \cite{doi:10.1080/01621459.1958.10501452}:
given the number of events $d_i$ occurring at time $t_i$ and the number of individuals surviving up to $t_i$ denoted as $n_i$, the survival function is estimated as $\hat{S}(t) = \prod_{i:t_i \le t} (1 - \nicefrac{d_i}{n_i})$.

In our case, we will measure line ``survival'' of code~\cite{gitoftheseus} introduced by ChatGPT and developers.
The lifetime of line starts with a commit that introduced it and ends at commit that no longer has it (this denotes specific content of a line). 
This is computed using reverse blame
(i.e. \texttt{git blame -{}-reverse} for each changed file,
limited to changed lines with sets of \texttt{-L} options).

%% file: 3-results.tex
\section{Analysis and results}
\label{sec:results}

\begin{figure*}[p]
  \centering
  \includegraphics[width=0.915\textwidth]{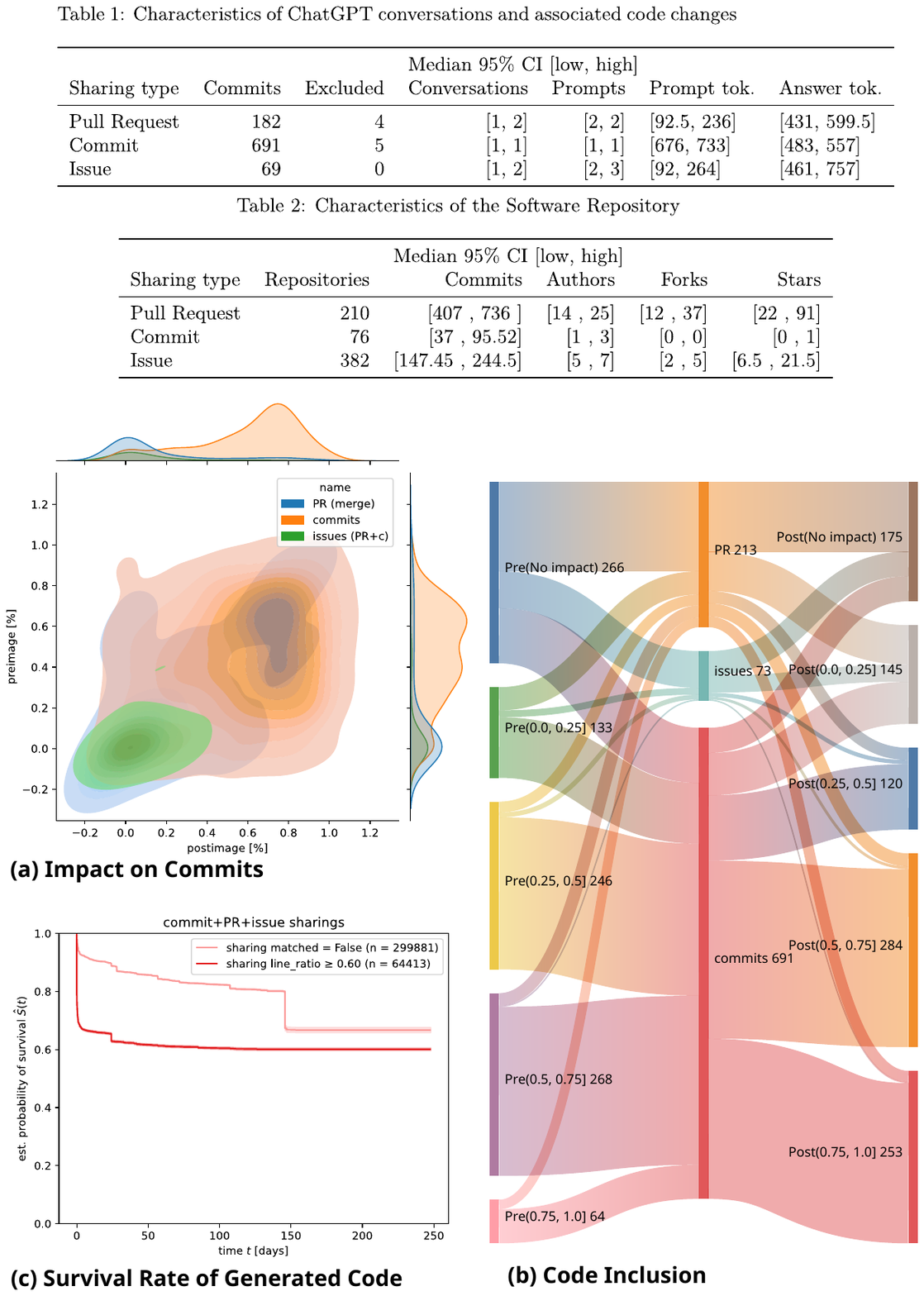}
  \caption{The data analysis.}\label{fig1}
\end{figure*}

{\bf RQ1.} To what extent does the usage of ChatGPT vary across different contexts of application, such as a commit (\DC), 
a pull request (\DP), or an issue (\DI)?

\textbf{Characteristics of conversations}.
In Tab.~1, for \DI, both median conversation and prompt counts fall within $[1,1]$ with 95\% confidence interval (\CI{[1,1]}). Pull requests and issues show slightly higher numbers. Median total prompt tokens are higher for commits than the other categories. Median answer tokens are similar across all types.
Our hypothesis is that this discrepancy may arise because, for \DC, participants tend to directly copy existing source code as prompts (large prompts).
It facilitates generation of a code-ready response within a single conversation.
Conversely, for \DI\ and \DP, ChatGPT serves more as a tool for drafting ideas, making points in discussions, or conducting reviews.
It is indicated by presence of multiple conversations with notably shorter prompts.

\textbf{Characteristics of the software repository}.
In terms of repository statistics, a broader diversity is evident (see Tab.~2. for details).
For \DC, the observed medians of characteristics are rather typical for small, less popular projects developed by small teams. This is different for \DI\ and \DP, where we witness more sophisticated projects across various dimensions (extended history, higher number of authors, and interactions).
This is another indication that we are witnessing different types of applications of ChatGPT between \DC\ and cases in \DP\ and \DI. 
Our intuitive explanation is that in the case of \DC, the lower number of co-authors and repository popularity in terms of fork number (see Tab.~2. for Commit type) might be correlated with smaller, personal like projects.

\textbf{Impact on commits}.
Next, we compare the degree to which ChatGPT is influenced by existing code and how it influences the resulting changes.
We start by comparing the distributions of changes, as depicted in Fig.~\ref{fig1}(a), focusing on aligning lines with ChatGPT conversations. For each code change, we evaluate the proportion of lines in the preimage likely provided by the user to ChatGPT (Y-axis) and the percentage of changes in the postimage (X-axis) likely attributable to ChatGPT's actions. The detailed procedure is outlined in Section \ref{sec:methods}.
Upon visual inspection of the graph, noticeable differences in the data distributions become apparent. To confirm this observation, we conduct a two-sample Kolmogorov-Smirnov test for goodness of fit on the marginal distributions.
We test the null hypothesis that two samples were drawn from the same distribution, with confidence level of $95\%$.
We will reject the null hypothesis in favor of the alternative if $P< 0.05$.
Initially comparing \DI\ and \DP\ for pre and post images, we obtain respective p-values of $P=0.09$ and $P=0.45$.
Thus we cannot reject the null hypothesis in this case.
On the other hand, when we compare \DC\ to either \DP\ or \DI, the resulting p-values are $P<0.001.$
Consequently, we reject the null hypothesis, favoring the assertion that the data were not drawn from the same distribution.
Therefore, we got another argument supporting the observation of different use cases for \DC\ versus \DI\ and \DP.

In Fig. ~\ref{fig1}(b), we divide and organize the code changes into bins. Within each bin, we count the number of code changes in both the pre and post images that have successfully matched the respective segments of the ChatGPT conversation.
We categorize the bins as follows: ``No Impact'' - for cases where there were no matching lines, and the intervals (0, 0.25], (0.25, 0.5], (0.5, 0.75], (0.75, 1] successively.
On the left side of the plot, we can observe how much ChatGPT was inspired by existing code, and on the right side, we observe to what extent the final change was inspired by ChatGPT. 
According to previous observations, we indeed see a difference in the utilization of ChatGPT. In the case of \DC, conversations contain a significant number of cues from existing code, and the resulting commits are much more influenced by ChatGPT. In contrast, for \DI\ and \DP, situations prevail where there is either no direct similarity or the similarity is minimal in both the pre-image and post-image.
This suggests that ChatGPT is not used as a code generation tool but rather as a tool to support discussions.

{\bf RQ2.} What happens with code influenced by ChatGPT in the repository? How does its lifetime compare to similar lines created by human developers?

\textbf{Survival Rate of Generated Code}
To analyze {\bf RQ2}, we conducted a survival analysis of post image code lines inspired by ChatGPT compared to all post image observed lines within \DC, \DI, \DP.
Surprisingly, there is a noticeable difference in the lifespan of lines of these different types.
Code lines inspired by ChatGPT undergo changes more quickly compared to lines without such inspiration (see Fig.~\ref{fig1} (c)).
The observed phenomenon may be attributed to factors such as a lack of project context, requiring refinement to meet project requirements and coding standards.
Additionally, the exploratory nature of ChatGPT-generated code may also lead to frequent modifications as developers iterate on ideas.

%% file: 4-discussion.tex
\section{Threats to validity}
\label{sec:validity}

\begin{enumerate*}
\item {\bf Observability of ChatGPT usage in a project} is one of the biggest limitations of this research.
DevGPT dataset construction depend on the developers' diligence to include ChatGPT url within their work on Github poses a major constraint. This limitation introduces a {\it selection bias} limiting its scope to announced code generated by ChatGPT.

\item {\bf Incompleteness of data}: Not all conversations are preserved, and users sometimes entered incorrect URLs, causing some entries in DevGPT to lack data about the conversations. Additionally, not all conversations in DevGPT are up-to-date compared to those available on the internet (we have seen examples where only a portion of the conversation was accessible in the dataset, possibly due to changes in the conversation after data acquisition).

\item {\bf Ethical Considerations}: We refrain from handling any sensitive individual contributor data and do not engage in automated judgments to attribute characteristics to individuals.

\end{enumerate*}

%% file: 5-related.tex
\section{Related work}
\label{sec:related}

With the widespread implementation of artificial intelligence-driven code generation, the software engineering research community has become increasingly interested in this phenomenon. 
A number of studies has been conducted in this area. Two major AI tools used to generate code are GitHub Copilot \cite{githubcopilot} and ChatGPT \cite{brown2020language}. 

Vaithilingam et al. conducted a study comparing Copilot with a widely-used traditional Intellisense, surveying 24 programmers \cite{DBLP:conf/chi/Vaithilingam0G22}. The majority of participants expressed a preference for Copilot, appreciating its ability to save time on online searches and streamline the programming initiation process. However, they raised concerns about the code generated by Copilot being more challenging to comprehend, and when errors occurred due to Copilot, debugging became a daunting task.

Jaworski and Piotrkowski conducted a similar survey~\cite{jaworski2023study}, and their findings echoed those from \cite{DBLP:conf/chi/Vaithilingam0G22}. However, respondents also raised concerns about potential data leaks and security.

Nguyen and Nadi took another approach \cite{DBLP:conf/icse/Imai22}. 
They used automatic tools to assess the quality of generated code for a number of programming languages.
They measured correctness by testing and complexity by software measurement.
They found out that Copilot-generated code had low complexity.
However, the correctness varied depending on the programming language choses. 

Imai tested Copilot in pair programing \cite{DBLP:conf/msr/NguyenN22}.
She compared it against a control group composed solely of human programmers. 
She reported that Copilot facilitated adding more code increasing the productivity.
However, that code was frequently deleted proving its inferior quality.
The results are seminal to our detailed research on the survival of ChatGPT-generated code. 

Nascimento et al compared the results achieved by human software engineers with ChatGPT-base solutions \cite{DBLP:journals/corr/abs-2305-11837}.
They compared the code submitted to Leetcode with that generated by ChatGPT. 
The results showed that ChatGPT is superior to novice and medium-experience programmers in case of up to medium-level problems. 
The authors have not found evidence that ChatGPT could be better that power programmers.

Yilmaz et al assessed the possibility to use ChatGPT in programming learning \cite{YILMAZ2023100005} by surveying students. 
The conclusions were in line with \cite{DBLP:journals/corr/abs-2305-11837}.
However, the authors report that students got lazy and were unable to provide complete answers to assignments. 
Additional measures were postulated regarding inclusion of ChatGPT into the learning process. 

Numerous studies have been performed to verify the usefulness of AI-generated code in day-to-day software production.
They are based on subjective user surveys or more objective local experiments with human-written and AI-generated code.
In our study we use the large meticulous dataset DevGPT \cite{devgpt} that facilities impartial assessment of the AI-generated code survivalability.
To some extent we confirm results of \cite{DBLP:conf/msr/NguyenN22} for ChatGPT on a significantly larger dataset.

%% file: 6-conclusion.tex
\section{Conclusions}
\label{sec:conclusions}

In this study, we evaluate ChatGPT's overall impact on software project maintenance using the MSR'24 challenge dataset~\cite{devgpt}.
In summary, our analysis sought to comprehend the diverse usage of ChatGPT across various contexts and its influence on source code.

We identified statistically significant differences in how ChatGPT was utilized in commits compared to issues and pull requests.
Our intuitive explanation suggests that for commits, users likely copy existing code as large prompts for swift, code-ready responses.
However, in pull requests and issues, ChatGPT serves more as a tool for idea drafting and discussions.

Our survival analysis indicated that ChatGPT-inspired code lines undergo changes more rapidly than non-inspired lines.
This is likely attributed to the lack of project context in ChatGPT's output, necessitating frequent refinement.
Moreover, the exploratory nature of ChatGPT-generated code may further contribute to its inclination for frequent iterative modifications.

\newpage